# ECONOPHYSICAL APPROACHES FOR THE DIRECT FOREIGN INVESTMENTS

Anda GHEORGHIU[*], Ion SPÂNULESCU[*] and Anca GHEORGHIU[*]

***Abstract.*** *In this paper will be applied some principles and methods from econophysics in the case of the direct foreign investitions (D.F.I.), particularised for the Greenfield type, and mixed firms of trade and industrial production (Joint Ventures). To this aim will be used some similarities and parallelisms between the mentioned economic domains and some phenomena and processes from physics, especially from thermodynamics, solid state physics (the grow of crystals and thin policrystalline layers etc.), electromagnetism etc.*

## 1. Introduction

Econophysics – science recently appeared – developed especially owed to the interpretations and models created for applications in the financial and investment domains, generally in the domains of industrial economy or goods consumption. Less has been the attempts to find the applicable models of the econophysics in the domains of the trade and marketing and especially in the trade and direct foreign investments domains.

By his nature of exact science which studies the laws of the nature, the physics proved to be the source and support of other sciences and especially of the technical ones (electrotechnics, metallurgy, electronics, civil constructions, machine engineering, thermotechnics etc.) but also for many boundary sciences as biophysics, geophysics, physico-mathematics, biomedical physics and more recenty econophysics and sociophysics.

Solid state physics, deals with the study and applications of the solid materials, especially metals, semiconductors and dielectrics with numerous and extremly important applications in electronics and microelectronics, in optics, in energetics (solar cells) etc., as well as with the plastic materials, polymers etc.

---

[*] Hyperion University, 169 Calea Călăraşilor, St., Bucharest-Romania

For the most diverse applications, solid materials (metals, semiconductors, oxides etc.), are used like thin films made by various methods from which the most used is that of evaporation and condensation in vacuum on an adequate solid support (substrate), especially in microelectronics and nanotechnologies.

Between the process of obtaining of thin films by evaporation and condensation from the gaseous phase and the process of the direct investments in the variants „greenfield", fusions, acquisitions or mixed firm associations (joint ventures), there are more similarities or analogies which shall be exposed and analised in this paper. The examination of these similarities between the two process types can lead to the settlement and/or understanding of some criterions or realisation conditions of some direct investments in other countries like „greenfield" investments on bare spots, or of fusions and acquisitions.

To be able to distinguish and understanding the similarities between the monocrystalline or policrystalline thin films obtaining process and the mode to realise the direct foreign investitions (D.F.I.), further on shall be succintly presented the conditions to obtain, and the main physical and structural characteristics of the thin films condensed in vacuum on plane supports warmed at various temperatures.

## 2. General considerations about the obtaining of the solid state thin films

To obtain the thin films by evaporation and condensation in vacuum some special devices are used: the evaporation precinct, in form of a glass case (Fig. 1) in which there are the evaporation source in form of a crucible containing the evaporation substance as well as the plane support 3 situated in the path of the atomo-molecular beam of the evaporated substance from the crucible. The support is warmed at various temperatures by dint of a separate oven (Fig. 1).

An important parameter to obtain a compact thin layer is the number of the incident particles on the support i.e. the atomo-molecular flux density produced by the evaporation source that must exced a critical value $I_c$ to have a condensation on the support. There is too a temperature $T_c$ of the support above which the condensation can't occur because $T$ support is too high [1].

The formation and growth of thin films process can be best explained by the adsorption theory of Frenkel and Langmuir and further developed by others researchers.

According to this theory the incident atoms on the support surface remain on it a time $t_s$ given by:

$$t_s = \tau \cdot e^{-\frac{E_d}{RT}} \qquad (1)$$

where $E_d$ is the adsorbtion (desorbtion) energy, and $\tau$ the molecular period of the adsorbed molecule (atom) vibrations ($\approx 10^{-14}$ s) on the support. During the time $t_s$ the adsorbed atoms migrate on the support similarity to bidimensional gas, after which they can reevaporate or to settle on the support.

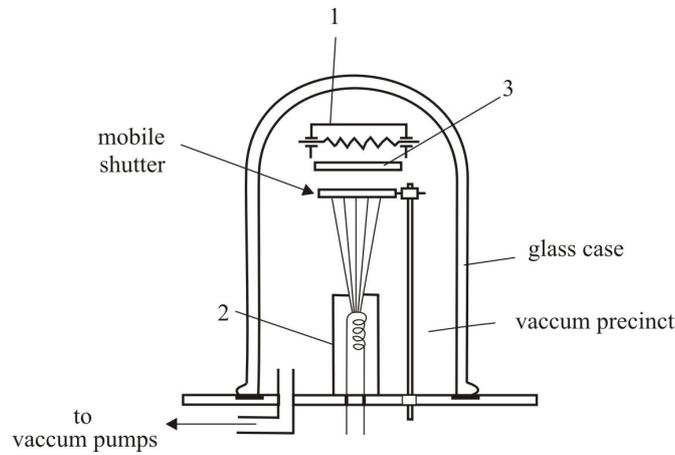

**Figure 1.** Vaccume evaporation equipment for thin solid films deposition.

The adsorbtion degree depends too on the accomodation coefficient α, defined by the probability of the incident atom (molecule) on the support's surface, to be adsorbed. The adsorbtion coefficients depends on the nature, temperature and surface support condition as well as on the nature and atoms (molecules) incident energy $E_i$. If $E_s$ and $T_s$ are the energy and the support temperature respectively, and $E_{st}$ and $T_{st}$ the energy and the layer temperature respectively, the accomodation coefficient may be expressed by [1.2]:

$$\alpha = \frac{T_{st} - T_i}{T_s - T_i} = \frac{E_{st} - E_i}{E_s - E_i}. \qquad (2)$$

In the case $T_s = T_{st}$ or $E_s = E_{st}$, then $\alpha \approx 1$ that corresponds, for instance, to the metal on metal condensation.

Numerous investigations showed that the nature and the state of the support deposition surface of the thin film presents a main importance for

27

the structure and the layer's physical properties. As function of the nature and support condition may be obtained monocrystalline layers (with some number of defects), policrystalline or even amorphous ones. It is supposed, of course, that the support temperature is high enough to obtain the crystalline phase, the amorphous one should appear in the case of low support temperature and the support surface contains oxides, traces of fatty acids etc. or an appreciable number of defects.

In the case of the supports with insufficiently cleaning or insufficient vacuum degassing or by other procedures, the obtained layers contain crystalls with a high number of structural defects.

It is generally admitted that the crystalline structure supports (ionic crystalls, semiconductors, metals, mica etc.) facilitate the apparition of cvasi-monocrystalline layers called epitaxial layers.

In case of autoepitaxy the support must be of the same nature and eventually with the same orientation and network type like the epitaxial layer (for instance, Si on Si, Ge on Ge, Cu on Cu etc.). It is supposed that all other conditions of epitaxial growth are realised (high support temperature, the absence of defects or impurities on the support surface, atom supersaturation of the deposed film etc.).

The amorphous supports like those of plane glass, generally lead to policrystalline or even amorphous layers.

## 3. Determination of the mass and the thickness of the condensed layers on the support

From the molecular physics it is known that inside a vacuum precinct containing a number $N$ of molecules with a total mass $M$, the number $v$ of molecules coming from all directions and hitting the surface unity in time unity is given by:

$$v = \frac{1}{4} n \bar{v} \qquad (3)$$

in which:

$$\bar{v} = 2 \left( \frac{2 k_B T}{\pi m} \right)^{1/2} \qquad (4)$$

is the arithmetic mean velocity of the molecules at temperature $T$.

By replacing the Boltzmann constant value $k_B$ and of the other numeric entities, the relation (4) becomes:

$$v = 4{,}66 \cdot 10^{24} \, p (MT)^{-1/2} \qquad (5)$$

in which *p* represents the gas pressure inside the precinct, and $M = mN$ is the molecular mass.

The adsorbtion velocity of the molecules deposed on the plane surface support placed on the way of the evaporation source from the inside of the precinct is given by:

$$w_{ad} = \frac{dn_a}{dt} = \alpha \nu \qquad (6)$$

where α is the accomodation coefficient, defined as the probability of the atom (molecule) incident on a surface to be adsorbed, namely fixed on the support by the Van der Vaals adhesion forces or by other forces of atomo-molecular nature.

Taking into consideration (5), the relation (6) becomes:

$$\frac{dn_a}{dt} = \alpha p (2mk_B T)^{-1/2} = 3{,}513 \cdot 10^{22} \alpha p \frac{1}{\sqrt{MT}}, \; [\text{cm}^{-2}, \text{s}^{-1}]. \qquad (7)$$

For the deposition of a layer of a substance of mass *m* it is necessary to establish a (dynamic) equilibrium between the solid phase (or liquid), from the support surface and the gaseous one (of the incident atomo-molecular flux). That is to say that the evaporation velocity $w_{ev}$ should be proportional to the mass *m* of the evaporated substance as well as to the adsorption velocity $w_{ad}$, of the particles from the incident atomo-molecular beam given by relation (7):

$$w = m \frac{dn_a}{dt} m \alpha_v p_v (2\pi m k_B T)^{-1/2}, \; \text{g/cm}^2 \cdot \text{s} \qquad (8)$$

where $\alpha_v$ is the evaporation coefficient (similar to the accomodation one α), and $p_v$ the saturated vapours pressure of the evaporated substance. Because $M = mN$, from (8) is obtained:

$$w = 0{,}0583 \alpha_v p_{v(\text{torr})} \sqrt{\frac{M}{T}}, \; \text{g/cm}^2 \cdot \text{s}. \qquad (9)$$

For the surfaces considered clean $\alpha_v = 1$.

From (8) and (9) formulas it is observed that the evaporation velocity depends on the substance nature (through the intermediary of the mass *M*), as well as on the thermodynamic evaporation temperature *T*.

Supposing a punctual evaporator with spherical simmetry, a sufficiently reduced pressure (to have an atomo-molecular beam) and a beam of not to high density (to neglect the shock number between the evaporated substance molecules), in order to calculate the condensated distribution on the receptacle support, the laws of Lambert from the geometric optics can be applied. From the optics it is known that the mean luminous intensity of a spherical punctual source, for the solid angle unity is:

$$I_0 = \frac{F}{4\pi}, \qquad (10)$$

where $F$ is the total flux emitted by the source in all directions. Analogously in the case of a punctual source from which a substance is evaporating in all directions with a constant velocity $w$[g/s], the substance quantity passing through a solid angle $d\Omega$ in any direction, in time unity is given by the relation:

$$dm = \frac{w}{4\pi} d\Omega. \qquad (11)$$

It is admitted that, after evaporation, the material is condensed on an elementary surface $dS_2$ whose normal forms the angle $\theta$ with the mean beam direction (Fig. 2).

Because the solid angle is given by:

$$d\Omega = \frac{dS}{r^2} \cos\theta, \qquad (12)$$

it results that:

$$dm = \frac{w}{4\pi} \frac{\cos\theta}{r^2} dS_2. \qquad (13)$$

In the case of a plane source emitting in the directions above the respective plane only (Fig. 2) the relation (10) it is written:

$$I_0 = F/\pi, \qquad (10')$$

and consequently the formula (18) becomes:

$$dm = \frac{w}{\pi} \frac{\cos\theta}{r^2} dS_2. \qquad (14)$$

If $\rho$ [g/cm$^3$] is the condensed substance density, then the thickness $d$, of the layer deposited in time unity can be found from the relation:

$$dm = \rho dV = \rho d \cdot dS_2. \qquad (15)$$

From (14) and (15) results for the thickness $d$:

$$d = \frac{w}{\pi\rho} \frac{\cos\theta}{r^2}. \qquad (16)$$

If the evaporation is from a punctiform source and the condensation is on a plane receiver surface $S$ (Fig. 2), then the layer thickness condensed in the point $P$ will be:

$$d = \frac{w}{\pi\rho}\frac{\cos\theta}{r^2} = \frac{w}{\pi}\frac{h}{r^3} = \frac{wh}{\pi\rho(h^2+x^2)^{3/2}}. \qquad (17)$$

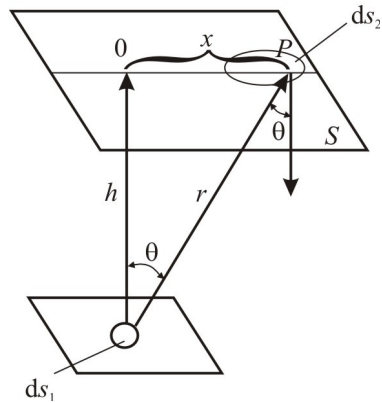

**Figure 2.** For calculation of the thin films thickness.

The layer thickness in the point 0 ($x = 0$) is given by (see rel. (17)):

$$d_0 = \frac{wh}{\pi\rho}\cdot\frac{1}{h^3} = \frac{w}{\pi\rho}\cdot\frac{1}{h^2}. \qquad (18)$$

In the thin films technology, in order to increase the productivity, inside the evaporation precinct can be situated more supports (Fig. 3,a), so from the source of mass $M$, $j$ layers can condensate of masses $m_1, m_2, m_3 \ldots m_j$, as it is seen from the figure 3,b.

Obviously, the thicher layers will be obtained on the supports placed just above the source $M$, at the minimum distance $h$ (Fig. 3) according to the relation (18), whereas the thickness of the other layers should be determined in compliance with (17) relation in which $x_j$ represents the distance from the point 0 (placed above the source) up to the considered support.

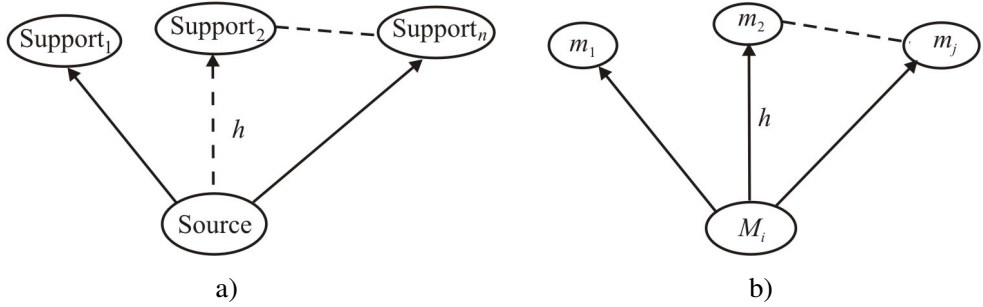

a)  b)

**Figure 3.** Deposition of several thin films from a single source.

It is to be mentioned that for the realisation of some compound substances layers or for the controlled impurification as in the case of semiconductor thin films that must be impurified, can be used many evaporation sources with $M_1, M_2 \ldots M_n$ mass tagether with many supports on which can be condensed layers of $m_1, m_2 \ldots m_p$ mass as can be seen in figure 4.

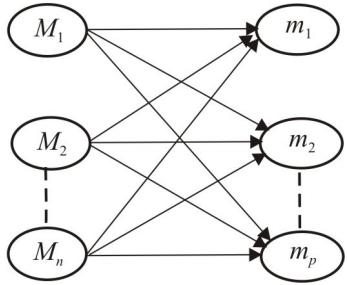

**Figure 4.** Deposition of several thin films from several evaporation sources.

Between the layer mass $m_j$ and the source mass $M_i$ there is a proportionality relation:

$$m_j = AM_i \qquad (19)$$

which in the general case with many sources and many condensation supports cand be written under the form:

$$\left.\begin{aligned}
m_1 &= k_{11}M_1 + k_{12}M_2 + \ldots + k_{1p}M_p \\
m_2 &= k_{21}M_1 + k_{22}M_2 + \ldots + k_{2p}M_p \\
m_3 &= k_{31}M_1 + k_{32}M_2 + \ldots + k_{3p}M_p \\
&\vdots \\
m_n &= k_{n1}M_1 + k_{n2}M_2 + \ldots + k_{np}M_p
\end{aligned}\right\} \qquad (20)$$

or under the form of a functional relation of the type:

$$\left.\begin{aligned} m_1 &= f_1(M_1, M_2, M_3 \ldots M_p) \\ m_2 &= f_2(M_1, M_2, M_3 \ldots M_p) \\ &\vdots \\ m_n &= f_n(M_1, M_2, M_3 \ldots M_p) \end{aligned}\right\}. \tag{21}$$

Coming again, for simplicity, to the case when we have a single source $M$ and a single plane, finite and regular (circular or rectangular) support placed above the source at a distance $h$ (Fig. 3), the mass $m_j$ of the deposed layer is direct proportional to the thickness $d_o$ of this one, i.e. (see rel. (18)):

$$m_j = kd_o = k\frac{w}{\pi\rho}\frac{1}{h^2} \tag{22}$$

in which it was taken in consideration the evaporation above the plane only, where is situated the evaporation source (Fig. 3). From the relation (22) it is seen that the deposed mass layer depends directly proportional to the evaporation velocity $w$ and inversely proportional to the distance support – source $h$.

## 4. The crystalline thin films growth model for the direct foreign investments analysis.

As it was mentioned in § 2, between the thin films obtaining process and the mechanisms of realisation of the direct foreign investments there are more similarities or analogies. Their examination may lead to the establishment or understanding of some specific criterions or realisation conditions of some direct investments types in other countries, markets or regions.

So, the mother-firm, with the economic-financial power (capacity) source $M$, can be assimililated with the source of mass $M_j$ from which is evaporating the substance that constitutes the capital fluxes guided towards various locations $j$ that can be assimilated with the support or with supports of masses $m_j$, $j = 1, 2 \ldots p$ placed in various locations (countries) $j$, at „economic" distances $h_j$ toward the source. By economic distance is understood the physical route (way) type as well as the infrastructure and the costs tied to the necessary utilities to realise the capital or equipments, raw materials etc., that can be assimilated with the "trajectory" of the atomo-molecular beam from the vacuum precinct in the case of thin films deposition.

In the case of the direct foreign investments (D.F.I.) the potential investments benefice $V_{ij}$ or the efficiency of the direct investments from the mass central source (the amount of $M_i$ funds of the mother firm), towards her branches or in other investments locations in various countries or regions characterised by "masses" or funds $m_j$, $j = 1, 2, 3 \ldots p$ can be calculated with the general relation:

$$V_{ij} = R_{ij} - C_{ij}. \qquad (23)$$

In (23) relation, $R_{ij}$ term represents the actual total value of the incomes expected to be obtained, and $C_{ij}$ is the actual value of the necessary costs to realise the investments $m_j$ in the location (or locations) $j$ (see Figs. 3 and 4).

The mass $m_j$ of the investments realised in the location $j$ characterises the value represented by density $\rho$ (i.e. the receptor region characteristic, from the point of view of the attractivity degree, permisivity for investments), as well as the volume $V_j$ of the investments (the value of each branch) so that:

$$m_j = \rho V_j. \qquad (24)$$

On the other hand, as it was shown (see rel. (22)) that "the mass" $m_j$ i.e. the value of the investments realised in location $j$ from other country or region, is directly proportional with the layer thickness $d_o$ i.e. with the dimensions or the volume of the realised investments:

$$m_j = k d_o = k \frac{w}{\pi \rho} \frac{1}{h^2} \qquad (25)$$

where $h$ represents the "economic distance" between the mother-firm (source) and the investments (branch) realised on a "plane" support at the location $j$, and $w$ is the transfer velocity (similar to "evaporation velocity") of the investments funds, equipments and necessary management.

But, as it was mentioned in the previous paragraph, the volume and the value of the realised investments ($m_j$ mass) are conditioned also by the value, the economic-financial power respectively and other characteristics of the mother-firm (source), fact that can be written under the form of the direct dependency (see rel. (19)):

$$m_j = AM_i. \qquad (26)$$

In the relation (26) *A* represents a specific parameter (proportionality constant) to each direct foreign investment type. *A* represents a managerial decision component at the mother-firm-level establishing the percentage from the total investment fund for a certain target. From (25) and (26) results the necessary value of the transfer velocity of the capital fluxes to location *j*:

$$w = \frac{\pi \rho}{k} AM_i h^2 = A_o M_i h^2. \qquad (27)$$

From the formulae (27) results that the capital flux transfer velocity, equipments etc. from the sourse *i* to the location *j* depends directly on the economical (capacity) power $M_i$ of the mother-firm and on the square of the economic distance from the source to the investment location. The remoter the respective location, the higher must be the value and transfer velocity of capital flux to compensate the supplementary expense with the transports that are conditioning also the realised products distribuition at the location *j* and the delivery of products respectively etc. It may be mentioned, for instance, the case of Nokia firm investments at the "support" industrial park Jucu near Cluj-Napoca from Romania. The selection of the respective locality being in large measure decided because the route distance *h* is smaller to the great towns from the nordic and central regions of Europe where must be distribuited the products and where from can come the equipments and installations (from Nokia "source") towards the location *j*, Jucu.

It is proved so that in the case of greenfield investments Nokia from the support Jucu near Cluj-Napoca, has taken in consideration the main conditions and exigences imposed by the realisation of some "layers" which constitutes the investments of mass $m_j$ and namely by the necessary "vacuum" in the work "precinct" characterised by the specific nature of the "support" (region) *j* having an accomodation coeficient or condensation α (see rel. (9)), symbolised by the cheap manpower attraction and disponible in the region, good conditions to bild the infrastructure (cheap and enough ground plot, construction materials etc.), the distance *h* relatively small between the source and "support" (*j* location) and the economic-financial power (capacity) $M_i$ of the mother-firm source as it results from the analyes of (27) formula too.

35

Similar considerations may be done also for other direct foreign investments categories for which can be detaled the differences of investments type, by the conditions and relations existing between the "source" and "condensation support". This can be done by the introduction of some suplementary, or correction factors, in $(22) \div (27)$ formulas and the adaptation of the ratiocination tied to the mother-firm characteristics and support respectively.

So, a joint venture supposes the previous existance in the target country one of the same nature firm (support), eventually of a reduced economic "power" on which a new patrimonial entity is in common realised, a profitable one.

In this case the joint venture investment type can be modell under the form of epitaxial growth[1] of some monocrystalline thin films on a massive monocrystalline support as it is, for instance, the growth of a Si layer on a Si – support or Ge on Ge, Cu on Cu etc. (Fig. 5).

Such mono or policrystalline thin films, named epitaxial, one can be obtained by substance transport from the source $S$ up to the support of the some nature like the deposed layer.

The transport of the gaseous substance can be done by dint of on inert gas or, easier, by deposition in vacuum precincts on supports (locations $j$) for which are valid the considerations and the precedent relations, especially $(22) \div (27)$ relations.

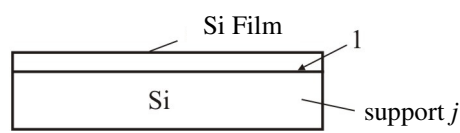

**Figure 5.** Grow of the epitaxial Si-film on the Si-support.

In figure 5, by 1 was marked the connection plane (joint venture) between the two regions (firms) that even have the same nature and crystalline structure etc., can have also some distinguishable characteristics, in the firm case, that from locality $j$ (the film deposed on $j$ support) may contain a more diversity of products of the firm $i$, as in the case of the semiconductor thin films that can have an other electronic conduction type, or other characteristics or optical or mechanical properties etc. slightly different from those of the mother-firm or of the main firm (source *i*) [1].

---

[1] The epitaxy term is coming from the greek words „επι" meaning „on" and „ταξις" meaning deposing on the same nature support, and was proposed for the denomination of the oriented crystal growts – one over the other, process.

## 5. Conclusions

In this paper it have been applied some principles and econophysics methods to model some direct foreign investments categories specially for greenfield and joint venture investments types. For that, have been used some similarities and analogies between the mentioned investments types and some processes or phenomena from physics, specially from thermodynamics and the solid state physics, with special reference to the crystalline or policrystalline substances growth, by their deposition under vacuum as thin films (layers).

In the paper it is shown that the thickness (as well as the volume) and the thin film deposed mass by vacuum thermal evaporation is determined by the evaporation velocity, i.e. by the evaporator temperature (source), by the mass $M_i$ and by the nature (density ρ) of the evaporating substance, as well as by the distance *h* between the source and the support on which the solid layer $m_j$ is deposed.

By dint of the proposed econophysic model it is established that between the obtaining of the thin films process and the realisation and analysis mechanisms of the direct foreign investments, may be identified more similarities or parallelisms (analogies), the examination of these being able to lead to the establishment or foundation of some specific criteria or realisation condition of some direct foreign investments types.

So, the mother-firm of economico-financial capacity $M_i$ can be assimilated with the "source" of $M_i$ mass from which is evaporating the substance that is constitute of the capital fluxes (similar to the molecular-beam from the solide substances evaporated in vacuum), directed towards the various locations *j* that can be assimilated with the support or supports of $m_j$, $j = 1, 2 \ldots p$ situated at various locations (countries) at "economic" distances $h_j$ with respect to the source.

In the paper is shown that the mass $m_j$ of the deposed layer, assimilated with the power or economico-financial capacity of the realised investment into the country *j* is proportional with the mass $M_j$ of source firm. The transfer velocity (similar to the evaporation velocity *w*) of the capital flux from the source $M_i$ to the location *j* depends directly on the economico-financial power $M_j$ of the mother-firm and from the square of the "economic" distance $h$ from the source to the investments location. In other words, the more far is the respective location, the larger must be the volume or the size of capital fluxes devilered by the mother-firm (investmental source) especially for the greenfiled investments types.

For the joint ventures investments type the grow process of thin epitaxial films model can be used, i.e. of a layer realisation (the investments from the location *j*) on a support of the **same nature** with that of the deposed material delivered (supplied) by the mother-firm that constitutes the 'evaporation" source.

The theoretical considerations regarding the value (size) of the capital fluxes from the source to the locations *j* are the same to those above mentioned for the greenfield investments type.

## REFERENCES

[1] I. Spânulescu, *Physics of Thin Films and Their Application* (in Romanian), Editura Ştiinţifică, Bucureşti, 1975.
[2] K. D. Chopra, *Thin Film Phenomena*, McGraw-Hill, New York, 1969.
[3] I. I. Frenkel, *Z. Physik*, *16*, 117 (1924).
[4] L. S. Palatnik, I. I. Papirov, *Epitaxialnîe plenki*, Nauka, Moskva, 1971.